%% file: sample-sigconf-authordraft.tex
\documentclass[sigconf]{acmart}

\usepackage{booktabs} % For formal tables

\setcopyright{rightsretained}
\usepackage{graphicx}
\usepackage{multirow}
\usepackage{subcaption}

\acmConference[KDD 2018 Workshop, AI for Fashion]{}{August 2018}{London, UK}
\copyrightyear{2018}
\begin{document}
\title{Understanding Fashionability: What drives sales of a style?}
%\titlenote{Produces the permission block, and
%  copyright information}
%\subtitle{What drives sales of a style?}
%\subtitlenote{The full version of the author's guide is available as
%  \texttt{acmart.pdf} document}

\author{Aniket Jain}
%\authornote{}
%\orcid{1234-5678-9012}
\affiliation{%
  \institution{Myntra Designs}
  %\streetaddress{}
  \city{Bangalore}
  \state{India}
  \postcode{560068}
}
\email{aniket.jain@myntra.com}

\author{Yadunath Gupta}
%\authornote{}
\affiliation{%
  \institution{Myntra Designs}
  %\streetaddress{}
  \city{Bangalore}
  \state{India}
  %\postcode{}
}
\email{yadunath.gupta@myntra.com}

\author{Pawan Kumar Singh}
%\authornote{}
\affiliation{%
  \institution{Myntra Designs}
%  \streetaddress{}
  \city{Bangalore}
  \country{India}}
\email{pawan.ks@myntra.com}

\author{Aruna Rajan}
%\authornote{}
\affiliation{%
  \institution{Myntra Designs}
%  \streetaddress{}
  \city{Bangalore}
  \country{India}}
\email{aruna.rajan@myntra.com}

%\author{Lawrence P. Leipuner}
%\affiliation{
%  \institution{Brookhaven Laboratories}
%  \streetaddress{P.O. Box 5000}}
%\email{lleipuner@researchlabs.org}
%
%\author{Sean Fogarty}
%\affiliation{%
%  \institution{NASA Ames Research Center}
%  \city{Moffett Field}
%  \state{California}
%  \postcode{94035}}
%\email{fogartys@amesres.org}
%
%\author{Charles Palmer}
%\affiliation{%
%  \institution{Palmer Research Laboratories}
%  \streetaddress{8600 Datapoint Drive}
%  \city{San Antonio}
%  \state{Texas}
%  \postcode{78229}}
%\email{cpalmer@prl.com}
%
%\author{John Smith}
%\affiliation{\institution{The Th{\o}rv{\"a}ld Group}}
%\email{jsmith@affiliation.org}
%
%\author{Julius P.~Kumquat}
%\affiliation{\institution{The Kumquat Consortium}}
%\email{jpkumquat@consortium.net}

% The default list of authors is too long for headers.
%\renewcommand{\shortauthors}{B. Trovato et al.}

\begin{abstract}
We use customer demand data for fashion articles on Myntra, and derive a fashionability or style quotient, which represents customer demand for the stylistic content of a fashion article, decoupled with its commercials (price, offers, etc.). We demonstrate learning for assortment planning in fashion that would aim to keep a healthy mix of breadth and depth across various styles, and we show the relationship between a customer's perception of a style vs a merchandiser's catalogue of styles. We also backtest our method to calculate prediction errors in our style quotient and customer demand, and discuss various implications and findings.%\footnote{This is an abstract footnote}
\end{abstract}


\maketitle

\input{samplebody-conf}

\bibliographystyle{ACM-Reference-Format}
\bibliography{sample-bibliography}

\end{document}

%% file: samplebody-conf.tex
\section{Introduction}

A fashion merchandiser builds their inventory by taking several attributes into consideration, such as which fashion article types to carry (eg. women's tops, kidswear, jeans for all, etc), what they should stand for (premium vs bargain vs fast fashion), and thereby what are the associated design attributes (fabric type, print, details, etc). On the other hand, a customer has a certain emotional connect with fashion that determines what s/he wears, how they shop for fashion, and how they perceive a brand or a retailer. How does a fashion retailer successfully interpret her merchandise in the
customer's view? The answer to this question is the key to building a more relevant inventory, fulfilling changing customer demand, and cutting losses on the long tail of inventory. 

At Myntra, every month, about 30 million customers browse, search for, and purchase our collection of about $5*10^5$ articles that span a range of known big label brands, Myntra's in-house fashion brands, and a marketplace where several small and medium scale brands list on our platform. Hence, we collect rich data on customer demand as well as available fashion inventory. In this paper, we propose a \textit{"Style Quotient"} or the customer demand for a fashion product's (hereby referred as style) content, that is independent of its commercials (price,
discount applied, promotional offers, advertising and marketing spend, etc). In order to mirror demand closely out with a full demand picture that is dependent on commercials, and deriving a decoupled style quotient that we propose to use for assortment planning.

\section{Style Quotient}~\label{style_q_def}
When looking at demand data, the choice made by a customer is hard to interpret as solely a matter of preference for the content of the purchased article, as sales are driven by merchandising factors like discount, list views (shelf space allocated in online store), marketing, and promotions. In this work, we show how we infer the influence of stylistic content (such as brand, color, fabric, fit, length, prints) on customer purchases. Today, in the fashion industry Rate of Sales, ROS (Sales Quantity/days live) is used as a proxy for customer preference for stylistic content  and we argue that in a highly dynamic environment such as ecommerce, where flash sales, festival discounts, and marketing notifications drive up demand, such a metric is non-representative of ``true'' customer preferences.

\begin{figure}
\includegraphics[scale=0.4]{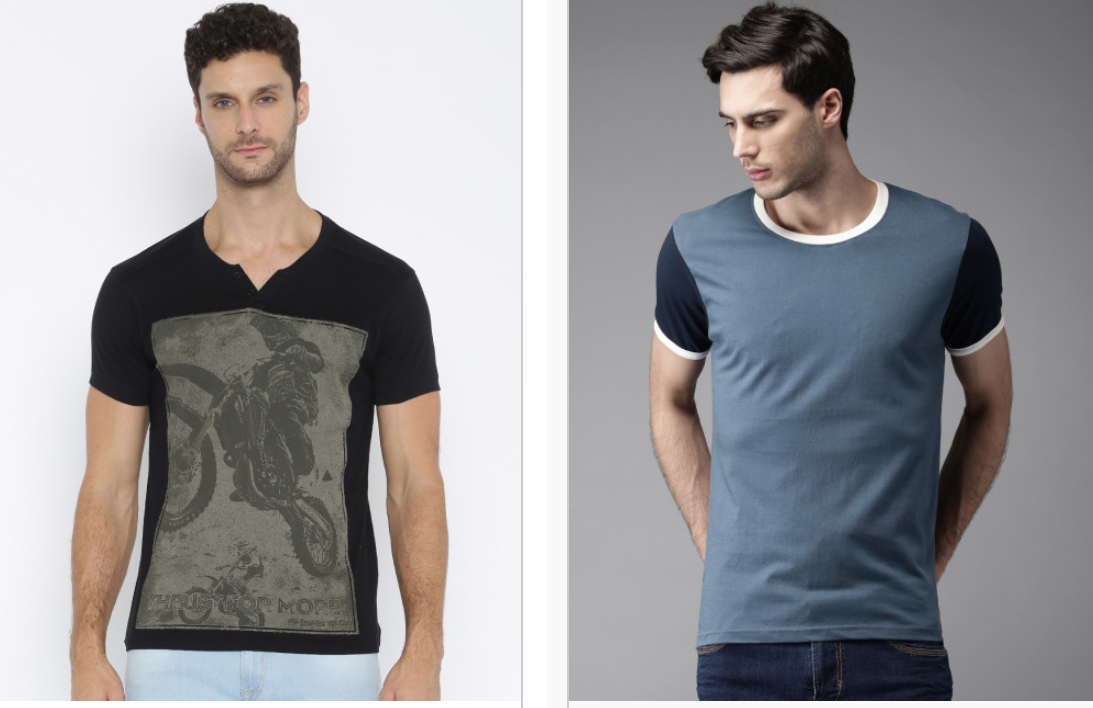}
\caption{Both styles have same ROS of 1.3. (a) \emph{Left} Style with high visibility (3,248 list views / day) and high discount (60\%). (b) \emph{Right} Style with low visibility (1,150 list views / day) and no discount (0\%).}
\label{style_trend}
\end{figure}

Figure \ref{style_trend} shows two fashion articles (styles) -- both with high ROS. Even though both styles have a similar ROS, style (b) is better than style (a) as its demand is less price and promotions driven. 
% article MRP, old vs new styles

% Bob, who is a buyer and part of procurement and planning team at Myntra, is interested in identifying top-selling styles that he should continue procuring, and bottom-selling styles that need efforts to be liquidated or de-listed from the portfolio. In order to assess the `goodness' of styles, Bob estimates Rate of Sales, ROS (Sales Quantity / Days Live) for each style based on recent sales data (last 1 month). However, ROS is coupled with merchandising factors. Figure \ref{style_trend} shows two styles -- both with high ROS. However, style on the left could drive high sales given high discount and visibility on the platform while style on the right achieved good sales regardless of any offers and low visibility. Clearly, ROS is a biased metric; style on the left needs to be liquidated as Bob needs to offer promotions to clear that inventory while style on the right should be procured as it sells on itself.}
% 

% \emph{Style Quotient}, SQ is a metric which quantifies the effect of stylistic content of a style based on customers' preferences in a given assortment. In this section, we will discuss the approach for estimating SQ. Intuitively, style on the right has higher SQ than on the left.

In order to compute the Style Quotient of a particular style listed on Myntra, we pick a subcategory of styles (article type - gender - elementary attributes based) within which comparisons are natural and easy to illustrate. A cross category style quotient can be computed by normalizing for width and depth appropriately, and without loss of generality, we will now deal with a subcategory alone in this paper. 
% At Myntra, we have a large catalogue of 530,000 styles spanning across multiple article types. This huge collection of styles is categorized into gender-article types -- Men-Tshirts, Men-Jeans, Women-Tops, Women-Dresses, etc. In our work, a given gender-article type is further divided in subcategories based on elementary attributes. For instance, `Men-Tshirts' is divided in subcategories based on fabric, sleeve length and collar. 
% An estimation of style quotient for each style in each subcategory for all article types demands an automated and scalable method. We now discuss our method of SQ estimation which is based on style sales data.  

%e.g. a higher list views or discount could imply a higher ROS. Bob can then end up buying styles that sell only with discount or he will not buy good styles that took a hit because of prioritization and higher list views of discounted styles. 

%Further, at Myntra, there is a large catalogue of 530,000 styles spanning across multiple article types, which further makes the problem of distinguishing good and bad styles even more difficult. Therefore, there is a need to design a metric that accurately captures the 'goodness of a style'. 

\subsection{Data}\label{sec:features}

We consider weekly sales / demand data for our computations. Intuitively, this may circumvent over-fitting due to frequent fluctuations in daily data, and under-fitting due to dissolution and averaging of driving factors in data at a coarser scale. The following attributes of a style are considered in modeling demand driven style quotient. A few  attributes are considered in their raw form and others are derived.

\begin{itemize}
\item \textbf{Raw}: \begin{itemize}
    \item \textit{Sales Quantity}: A numerical measure indicating the actual sales of the style.
 %   \item \textit{Product Gender}: A categorical measure signifying whether the product is designed for man, woman, kids or unisex. 
    
    \item \textit{Is Live}: A binary measure indicating whether the style is live on platform or not. 
    
    \item \textit{First Time on Discount}: A binary measure indicating if the style is put on discount for the first time. This may attract additional traction and increase sales. 
    
    \item \textit{Number of styles live from same brand}: A numerical measure quantifying the competing styles from the same brand. Many similar styles may lower sales of a particular style.

\end{itemize}

%The following derived features are also used:

\item \textbf{Derived}: \begin{itemize}
    \item \textit{Discount Deviation}: Dispersion around average selling price. 
    
    \item \textit{Normalised list price}:to indicate whether a style is premium or for the mass market. 
    
    \item \textit{List Views Deviation}: List view is the shelf space allocated to a style in an online store. List views deviation is a numerical measure of style visibility dispersion indicating if a style is attracting higher views as compared to an average for reasons like promotion; making the style noticeable early than others. 
    
    \item \textit{Style Age}: A measure indicating the shelf life of a style. With longer shelf life, the style's demand may decay with time. 

\end{itemize}

\end{itemize}

\subsection{Design}\label{sec:design}
We capture the customer preferences for an assortment using `demand prediction' framework. Let there be a universal set $S$ = \{$s_{1}$, $s_{2}$, $\cdots$, $s_{N}$\} which represents the store's all styles for a given subcategory present in observed time duration $T$. The store's assortment at week $t$ is represented by $A_{t}$ = \{$s_{i}$ $\in$ $S$ : $s_{i}$ live at $t$\}. 

Customers' preferences is captured as probability of choosing a style $s_{i}$ at week $t$ and is denoted by $p_{it}$. Customers choose a particular style based on style's content and merchandising factors such as discount, list views, MRP, and promotion present in week $t$. We use Multinomial Logit (MNL) model to derive customer preferences, where $p_{it}$ is given by (\ref{eq:mnl}). 

\begin{equation}
\label{eq:mnl}
p_{it} = \tfrac{\exp{(U_{it})}}{\sum_{\\{s_{j}} \in A_{t}}\exp{(U_{jt})}}  \qquad \qquad \qquad \forall s_{i} \in A_{t}
\end{equation} 

where $U_{it}$ is utility attached to style $s_{i}$. Style utility, $U_{it}$ is dependent on style's content and merchandising factors in week $t$. We use log-centered transformation on  (\ref{eq:mnl}) to estimate customer preferences (see ref \cite{kok2007demand} and \cite{cooper1988market}). 

\begin{equation}
\label{eq:demand}
U_{it} = \ln(\frac{p_{it}}{\bar{p}_{t}}) = \sum_{\\s_{j} \in S}\gamma_{j}I_{ij} + \sum_{k = 1}^{K}\beta_{k}(f_{ikt} - \bar{f}_{kt}) + \epsilon_{it} \qquad \forall s_{i} \in A_{t}
\end{equation} 

where $\bar{p_{t}}$ is the mean choice probability over all styles live at $t$, $I_{ij}$ = $\{1,$ if $i=j; 0$ otherwise\}, $\gamma_{i}$ is the style-specific effect for style $s_{i}$,  $f_{ikt}$ represents time-varying merchandising factor $k^{th}$ feature, $\bar{f}_{kt}$ is the mean of the $k^{th}$ feature in the subcategory and $\epsilon_{it}$ is the error term. We use sales data in order to empirically compute $p_{it}$ as the ratio of the number of customers who bought style $s_{i}$ to the number of customers who bought any product in $A_{t}$. We fit linear regression to estimate parameters $\gamma_{j}$ and $\beta_{k}$ using Least Squares method. 
Style Quotient, $(SQ)_{i}$ for style $s_{i}$ is derived based on style-specific effect as follows:
\begin{equation}
(SQ)_{i} = \exp{(\gamma_{i})}
\end{equation} \label{eq:sq}

We choose a parametric model to determine style quotient, as this metric determines a style's `fashionability', and is dependent on factors like look, quality, appeal which are subjective and difficult to quantify. Instead of computing it as a function of intangibles such as look and appeal, we propose its estimation as an additive contributor to customer choice over and above the merchandiser's promotions. 
%($p_{it}$). 

\section{Experiments}
We now demonstrate the usefulness of our construct, by predicting measurable outcomes using our style quotient. 

We consider 20,082 styles of men's tshirts spanning  5 subcategories bought over a period of 26 weeks. Subcategory details are shown in Table \ref{tab:subcategory}. We consider only those styles that were  listed for at least 4 weeks and construct the related feature set as explained in section~\ref{sec:features}. 

\begin{table}[h]
  \caption{Dataset: Styles in Men-Tshirts Subcategories}
  \label{tab:subcategory}
  \begin{tabular}{clr}
    \toprule
    Subcategory&Description&No. of Styles\\
    \midrule
    1 & Short Sleeves, Polo Collar & 5,179\\
    2 & Short Sleeves, Round Neck & 9,799\\
    3 & Short Sleeves, V-Neck & 2,012\\
    4 & Long Sleeves & 2,486\\
    5 & Sleeveless & 606\\
  \bottomrule
  & Total & 20,082\\ \bottomrule
\end{tabular}
\end{table}

%\subsection{Analysis and Results}
\subsection{Evaluation}
We use the data for first 22 weeks for each subcategory as training data and estimate sales for the next 4 weeks as part of test. Our baselines are -- \\ 
\begin{itemize}
\item
\textbf{Simple Rate of Sales (ROS)}: Sales for the test period is estimated based on average ROS of last 4 weeks during the training period. This metric captures the recent demand of the style. However, it does not capture the subcategory level sales trend. \\

\item 
\textbf{ Normalized Rate of Sales (ROS)}: Let, total sales for a given subcategory at time $t$ be $D_{t}$, and $d_{it}$ represents estimated sales for style $s_{i}$ within the subcategory. $d_{it}$ is computed as follows: 
\begin{equation}
    d_{it} = \tfrac{(ROS)_{i}}{\sum_{s_{j} \in A_{t}}(ROS)_{j}}*D_{t}
\end{equation}

\item
\textbf{Mean Intercept Demand Prediction}: We train a linear regression model using training data and a non-varying intercept; such model does not capture the style specific effects. Essentially, asssuming that each style has same `fashionability'. Mathematically, 
\begin{equation}
\ln(\frac{p_{it}}{\bar{p}_{t}}) = \beta_{0} + \sum_{k = 1}^{K}\beta_{k}(f_{ikt} - \bar{f}_{kt}) \qquad \forall s_{i} \in A_{t}
\end{equation} \label{eq:midemand}

$p_{it}$ is estimated using (\ref{eq:mnl}). $d_{it}$ is computed as follows: 
\begin{equation}
\label{eq:sales_pred}
    d_{it} = p_{it}*D_{t}
\end{equation}

\item
\textbf{Style Quotient Based Demand Prediction}: We capture the style specific effects in this model by replacing $\beta_{0}$ with style-specific effects $\gamma_{i}$ (see equation~\ref{eq:demand}). $d_{it}$ is estimated using (\ref{eq:sales_pred}). 
\end{itemize}

$D_{t}$ can be estimated using suitable models, but for the sake of comparison across different benchmarks we are using actual sales data. Sales prediction error is calculated using weighted-MAPE (wMAPE) i.e. the mean absolute deviation from actual sales. Mathematically, defined as: 
\[
wMAPE = \frac{\sum(|A-F|)}{\sum{A}}
\]

where, $A$ = Actual sales, $F$ = Predicted Sales. Lower the wMAPE, better is the prediction.

\subsection{Results}
Table~\ref{tab:SQ_perf} shows wMAPE with various baselines and SQ based prediction. With SQ in consideration for sale prediction, overall wMAPE significantly reduced by \textbf{20.9\%} over normalized ROS based prediction and by \textbf{10.6\%} over mean-intercept model. Reduced error thus imply that estimated style quotient helps in predicting the future sales better as compared to current methods.

%We train a linear regression model (refer Section~\ref{sec:design}) based on first 22 weeks data for each subcategory. Last 4 weeks data is used as test data. First, we discuss various qualitative measures to evaluate Style Quotient. Then, quantitative results are presented on Style Quotient via demand prediction. 

\begin{table}[h]
\small{
  \caption{Evaluation using wMAPE on test data across subcategories. Lower wMAPE and significant improvement with SQ based predictions than baselines. }
  \label{tab:SQ_perf}
  \begin{tabular}{p{0.8cm}ccp{0.8cm}ccc}
    \toprule

    \multirow{3}{*}{} & Simple & Normalized & Mean & SQ & \multicolumn{2}{c}{Improvement} \\
    Sub & ROS & ROS & Intercept & Model & (d) vs (b) & (d) vs (c)  \\
    category & (a) & (b) & (c) & (d) & & \\
    \midrule
    1 & 73.0 & 66.5 & 55.2 & 47.2 & 19.3 & {8.0}\\
    2 & 73.2 & 72.0 & 61.1 & 48.6 & 23.4 & {12.5}\\
    3 & 74.8 & 67.3 & 52.3 & 43.1 & 24.2 & {9.2}\\
    4 & 58.8 & 53.4 & 49.1 & 37.8 & 15.6 & 11.3\\
    5 & 58.3 & 53.5 & 44.9 & 40.8 & 12.7 & 4.1\\
  \bottomrule
  Overall & 70.3 & 66.4 & 56.1 & {45.5} & \textbf{20.9} & \textbf{10.6}\\ \bottomrule
\end{tabular}
}
\end{table}\label{tab:subcat}

Table~\ref{tab:SQ_week_perf} shows that projections based on ROS based models are highly inaccurate as we move further away from recent time (wMAPE ranges from 59.4\% to 85.2\% for Simple ROS; 55.9\% to 80.5\% for Normalized ROS) while regression models with or without style specific factor are stable and produce much less erroneous predictions as the models consider fluctuations of merchandising factors. Further, SQ based predictions are much less erroneous than mean intercept model predictions by ~10\%.

\begin{table}[h]
\small{
  \caption{Evaluation using wMAPE on test data over time. Lower and stable wMAPE for Mean Intercept and SQ based predictions than baselines. Merchandising factors and Style specific factors that vary with time helps in maintaining stable error rate. }
  \label{tab:SQ_week_perf}
  \begin{tabular}{cccp{0.8cm}ccc}
    \toprule
    \multirow{3}{*}{} & Simple & Normalized & Mean & SQ & \multicolumn{2}{c}{Improvement}  \\
     Week & ROS & ROS & Intercept & Model & (d) vs (b) & (d) vs (c)  \\
     & (a) & (b) & (c) & (d) & & \\
    \midrule
    23 & 59.4 & 55.9 & 56.8 & \textbf{45.4} & 10.5 & \textbf{11.4}\\
    24 & 63.9 & 62.4 & 53.9 & \textbf{44.1} & 18.3 & \textbf{9.8}\\
    25 & 73.7 & 67.4 & 55.2 & \textbf{45.2} & 22.2 & \textbf{10.0}\\
    26 & 85.2 & 80.5 & 58.6 & \textbf{47.5} & 33.0 & \textbf{11.1}\\
  \bottomrule
\end{tabular}
}
\end{table}\label{tab:week}

\begin{figure}[h]
    \includegraphics[scale=0.4]{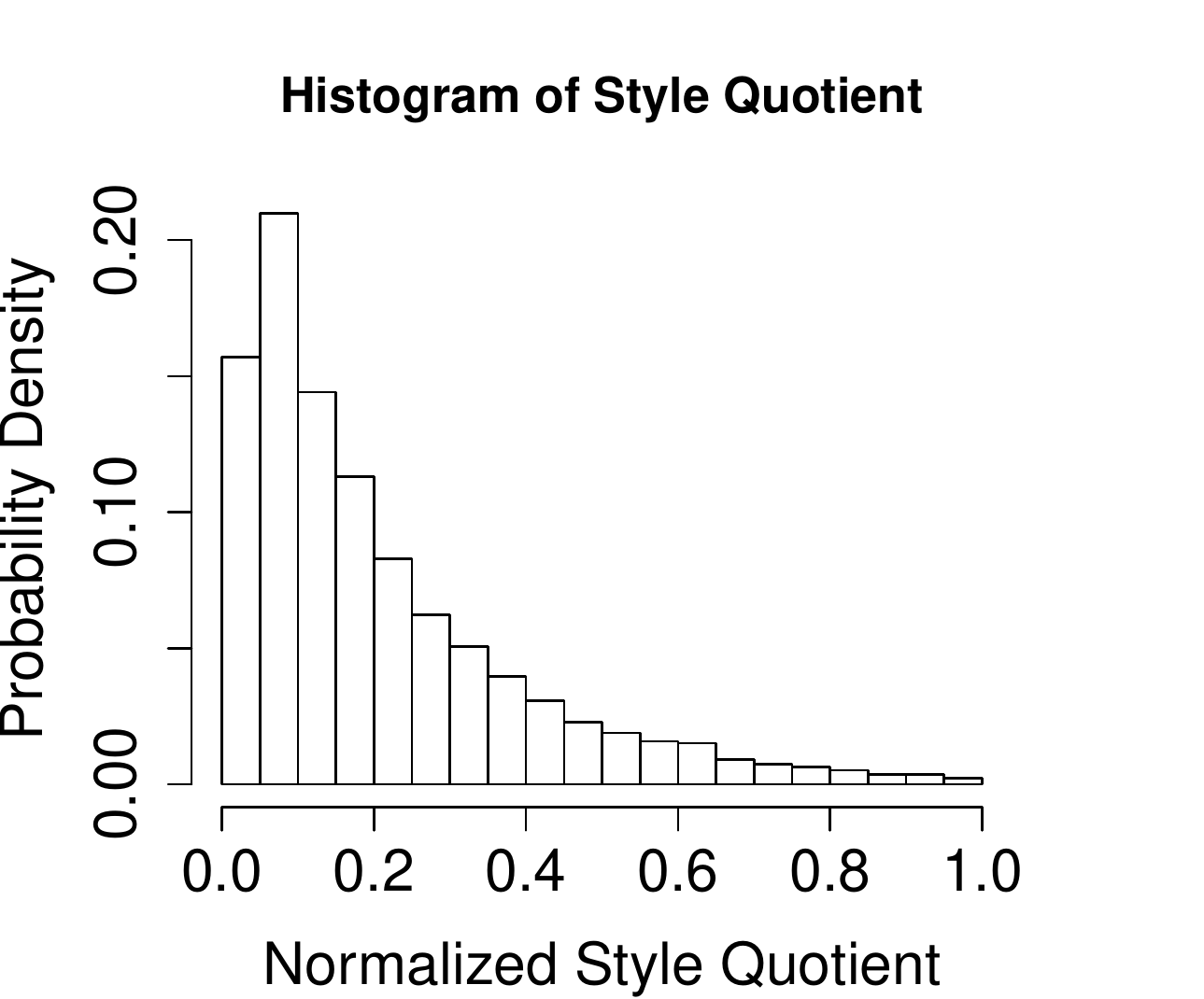}
    \caption{Histogram of Normalized Style Quotient; shows a positively skewed distribution with mean 0.21 and standard deviation 0.18. 15\% styles have high SQ (>0.4). These styles are highly fashionable.}
    \label{sq_hist}
    \end{figure}

\begin{figure*}[t]
     \begin{subfigure}[b]{0.3\textwidth}
    \includegraphics[width=\textwidth]{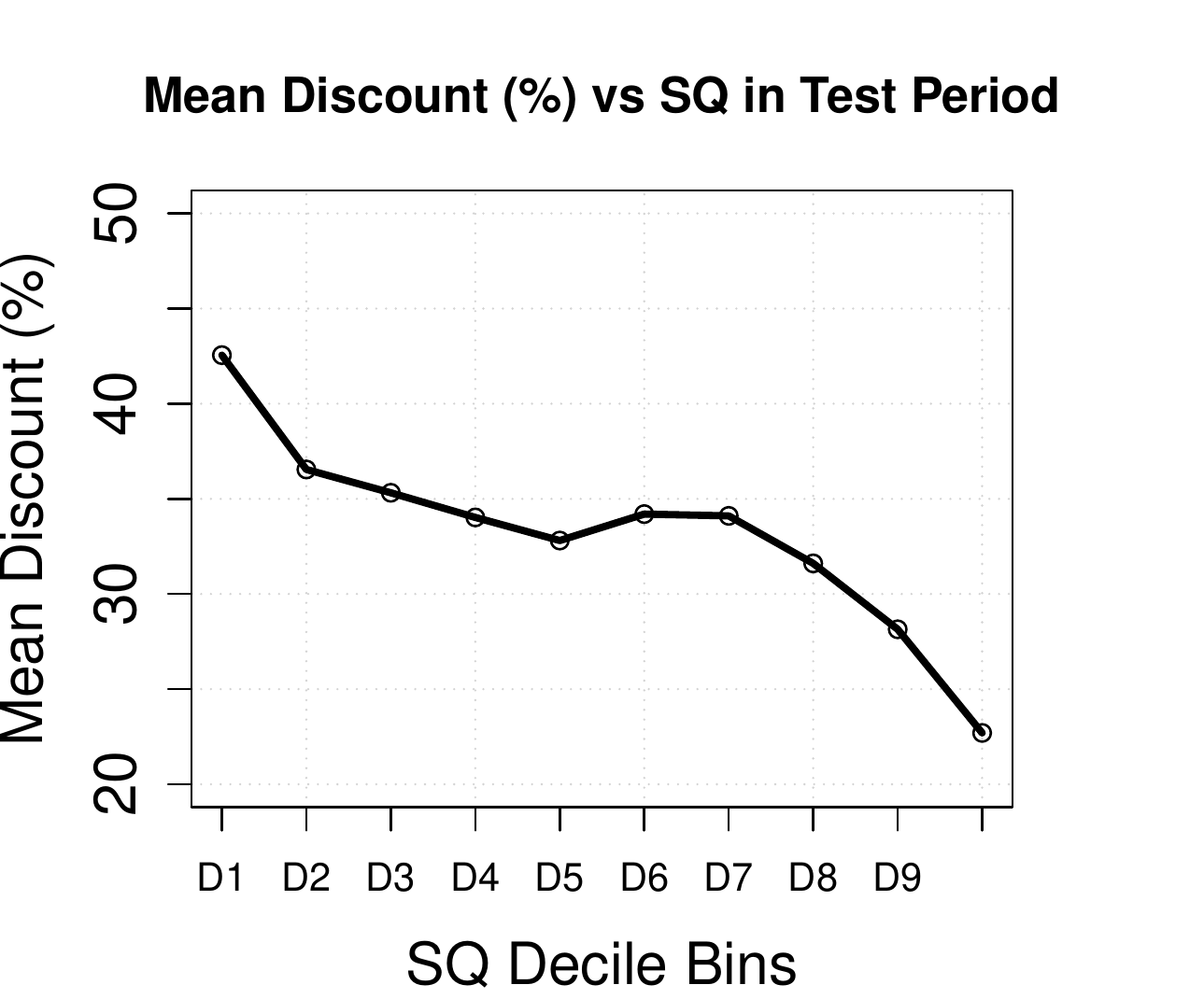}
    \caption{Discount variation with SQ.}
    \label{sq_discount}
    \end{subfigure}
\qquad
 \begin{subfigure}[b]{0.3\textwidth}
    \includegraphics[width=\textwidth]{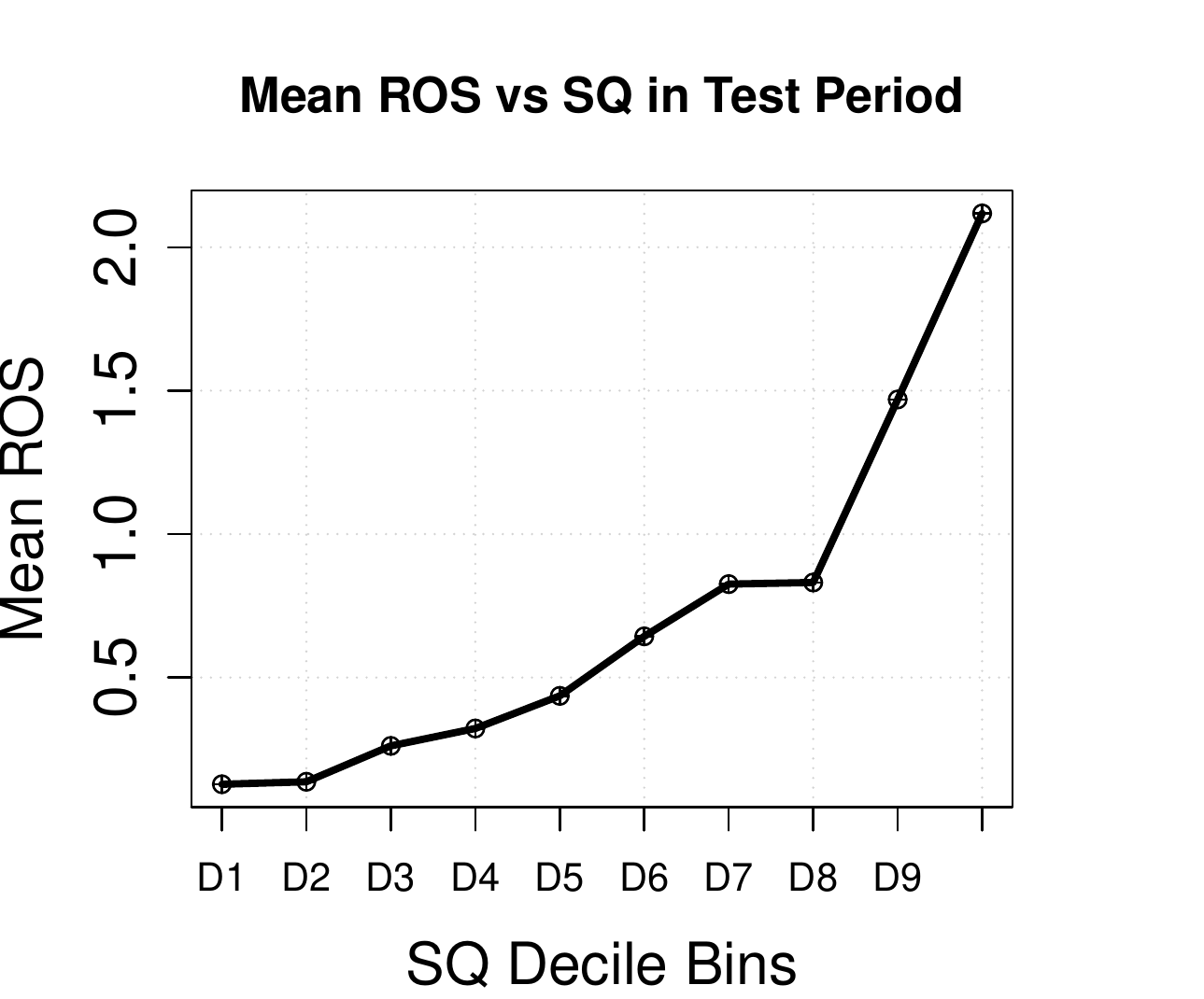}
    \caption{ROS variation with SQ.}
    \label{sq_ros}
    \end{subfigure}
    \caption{shows SQ variation with discount and ROS. Styles with high SQ sell on less discount and return more ROS while low SQ styles need more discount to attract customers and promote sales.} \label{sq_ROS_discount}
\end{figure*}

Thus, style quotient helps answer the questions -- a) We can better predict the sales with style quotient in consideration than without it; b) Style quotient captures `fashionability' which varies with style and capture style's intrinsic demand and appeal. 

\subsection{Qualitative Analysis}
In this section, we discuss interesting insights and properties derived for Style Quotient. We analyze Style Quotient for subcategory 2 without loss of generality. For this analysis we make 10 bins on SQ, based on deciles (D1 being lowest SQ bin and D10 being highest SQ bin), to analyze its relationship with other performance characteristics.  

\begin{figure}[htp]
\includegraphics[scale=0.4]{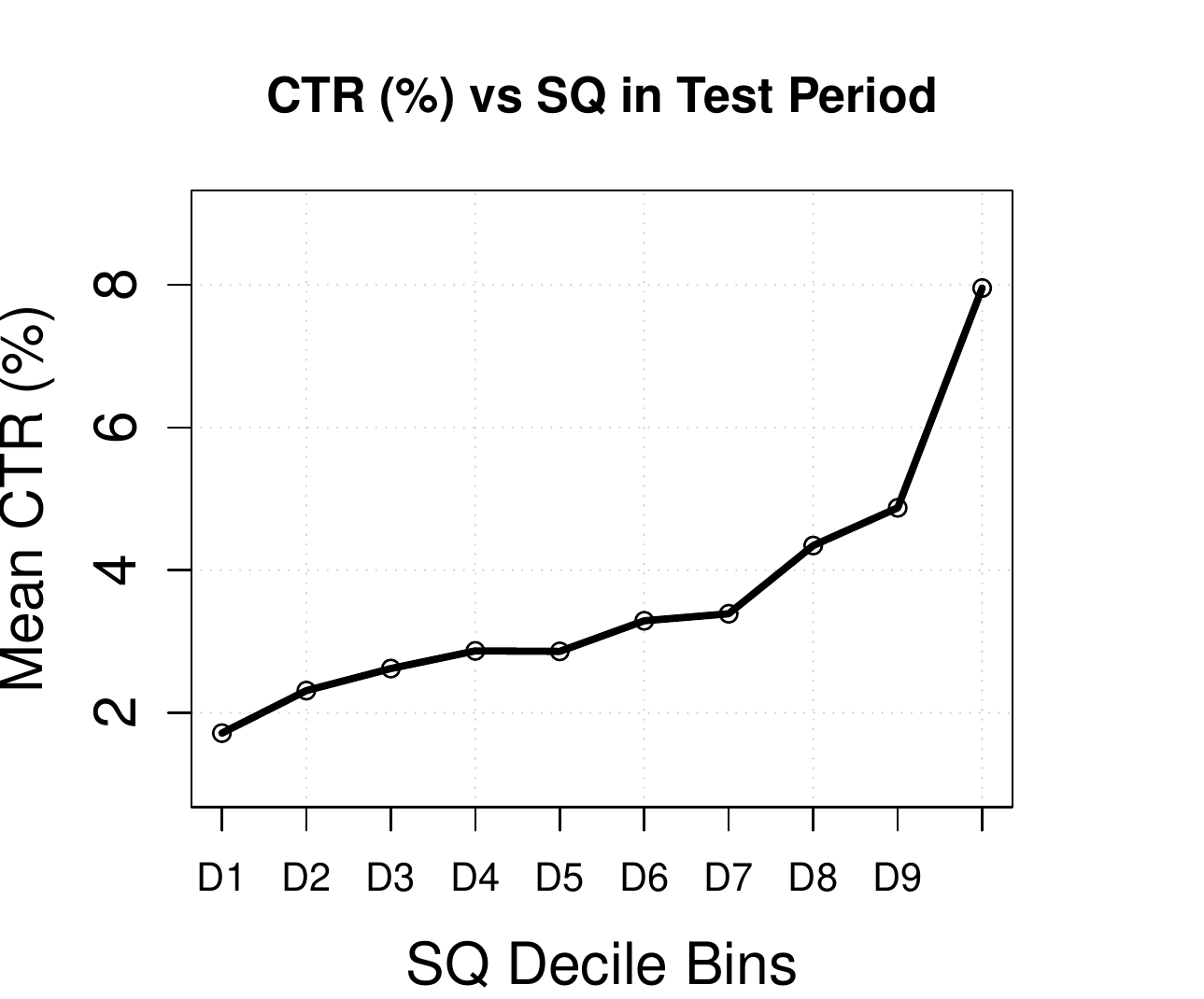}
\caption{CTR variation with SQ: shows higher customer interest for higher SQ styles}
\label{sq_ctr}
\end{figure}

\begin{itemize}
\item \textit{Style Quotient Distribution}: Figure~\ref{sq_hist} shows the positively skewed distribution of Style Quotient normalized between 0 to 1. This indicates few styles having high SQ, as expected in Fashion industry.

\item \textit{Discount and ROS variation with SQ}: As shown in Figure~\ref{sq_discount} and Figure~\ref{sq_ros}, discount decreases and ROS increases with increasing SQ. This indicates its easier to sell high SQ styles at low discount, compared to low SQ styles.

\item \textit{Click Through Rate (CTR) variation with SQ}: CTR is defined as the ratio of the number of times customer clicks on a product to the number of times product is shown. Figure~\ref{sq_ctr} shows increase in CTR with increasing SQ indicates higher customer interest for higher SQ styles. Thereby, indicating effectiveness of SQ in identifying better stylistic content which appeals to customers.  
        
\end{itemize}

\section{Style Quotients in Fashion Retail}
In this section, we discuss how to operationalize a fashion retail supply chain on the basis of our proposed SQ. 
\subsection{Top-Seller Identification}
From Figure~\ref{sq_ROS_discount}, it is clear that styles with high SQ sells at higher ROS and lower discounts. Thereby, indicating that styles with higher SQ are top-sellers. Hence, replenishment and planning must focus on higher SQ styles, to improve overall margin and assortment health. 

\subsection{Liquidation of Styles}

\begin{figure}[ht]
\includegraphics[scale=0.4]{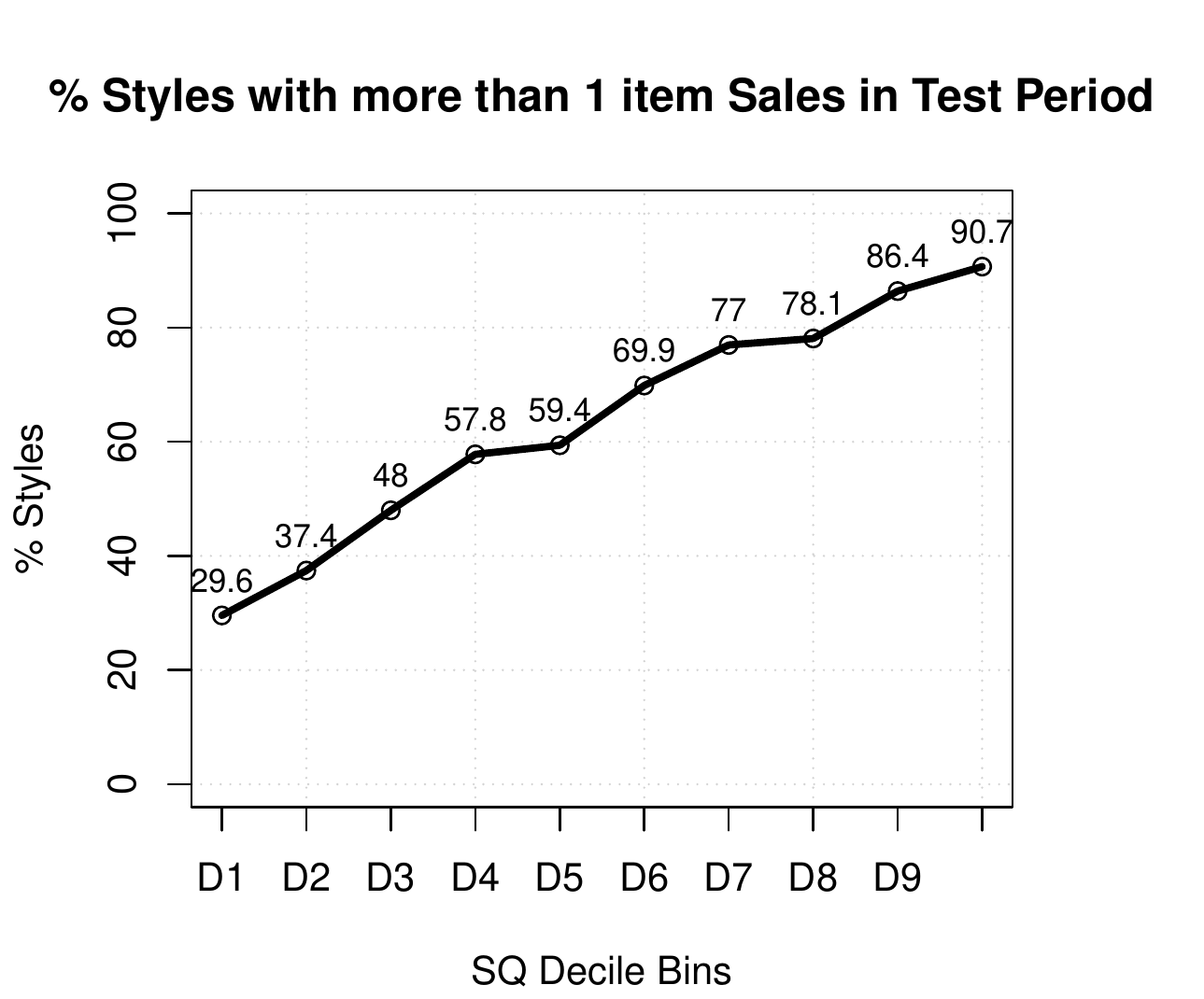}
\caption{shows lower SQ Bins have lower chances of Sales in future; These styles need to be liquidated.}
\label{liquidation}
\end{figure}

Styles with lower SQ are potential subset for liquidation, as the expected return on low SQ styles (given demand elasticity curves) is poorer than with higher SQ styles, given holding costs and margins. Figure~\ref{liquidation} and accompanying chart show a clear upward trend for sales ahead with increasing SQ. 

\subsection{Assortment Planning}
The key objective of any fashion retailer is to appeal to the fashion aesthetics and needs of the customer segment they serve, and hence any assortment planning activity must aim to increase the average style quotient of their inventory, and soften the long tail in inventory management. The longer tail (by frequency of sale) of products should preferably of higher SQs so that a true depth vs breadth optimization can be achieved in assortment planning. 

Table \ref{tab:sq_brand} shows the average SQ across various brands carried on Myntra. One way to increase average platform SQ could be to increase the representation of brands having higher mean SQ. 

\begin{table}[h]
    \centering
 \caption{Mean SQ across Brands on Myntra}
 \vspace{-3mm}
    \begin{tabular}{ccc}
         \toprule
         Brand & No. of Styles & Mean SQ \\
         \midrule
         B1 & 148 & 0.386 \\
         B2 & 175 & 0.335 \\
         B3 & 91 & 0.326 \\
         B4 & 270 & 0.324 \\
         B5 & 493 & 0.252 \\
         B6 & 54 & 0.103 \\
         B7 & 201 & 0.077 \\
         B8 & 103 & 0.069 \\
         \bottomrule
    \end{tabular}
    \vspace{-2mm}
    \label{tab:sq_brand}
\end{table}

% \subsection{Grading for New Styles} don't know of any way that SQ suggests this apart from looking at demand for content, which in this case is inferred not computed as a function of content. 

\section{Related Work}
Traditional approaches for retail assortment optimization are reviewed in great detail in \cite{kok2008assortment}. Most of these models are based on consumer choice models. In \cite{ryzin1999relationship}, Multinomial Logit (MNL) model is used to find the optimal inventory for assortment substitution in a category. This work is extended for stock-out substitution in \cite{mahajan2001stocking} and for further scenarios in \cite{cachon2005retail, maddah2007joint, caro2007dynamic}. In \cite{smith2000management, kok2007demand}, assortment planning problem is studied with Exogenous Demand model and an integer programming formulation. In \cite{gaur2006assortment} it is shown that the products in the optimal assortment are far apart and there is no substitution between products. These models assume that the customers have a clear preference for the product they want to buy. If the preferred product is not available in the assortment, then the customer may substitute a different product based on a well define substitution probability. However, most of these studies are related to hard-goods and grocery segments which have long life spans and minimal variation in customer preference. Whereas in fast fashion industry, products have a short life span and ever changing customers preference, with choices being influenced by merchandising factors like discounts, advertisements etc.

Apart from assortment optimization, stylistic content of products has also been discussed in the context of personalized recommendations. \cite{bracher2016fashion} has done some work in the direction of fashionability of articles by using style embeddings for product recommendation. Fashion articles are also ranked based on the likeability of their observable and latent visual features in \cite{wang2015towards}.  However, it does not account for merchandising factors which significantly affect demand and hence the assortment decision. A similar attempt to quantify stylistic content has been made in \cite{garg2016sales}, though it has limited applicability due to the constraint of visual similarity.

\section{Conclusion}
In this work, we propose an indirect metric to assess fashionability of fashion articles : Style Quotient (SQ). We calculate style quotients for our assortment mix, and show that this metric is more stable to variations in external parameters (to style) such as discount and visibility provided on the selling platform. It is, hence, a better predictor of future demand for a style, as well as a key metric to increase while building fashion brands. While a diverse range of style quotients are needed to support a healthier mix of assortment appealing to various tastes, the average style quotient is a good indicator of the inherent match between a fashion retailer's merchandise and its target segment's tastes. We use a reductionist inferencing approach that does not make any assumptions on content of a style in its appeal to the user, but rests solely on user behaviour observed on our site. 
The long tail problem in fashion needs clear quantification to study, track, and optimise for, and our work is a first fashion retail specific approach to the problem, to the best of our knowledge. The various applications we discuss are currently operationalised or in process at Myntra, and therefore we demonstrate practical usefulness in decision making of our work.

% \iffalse
% \subsubsection{Qualitative Analysis}

% As discussed in Section~\ref{style_q_def}, we now show that Style Quotient is an `un-biased' metric on merchandising factors as compared to ROS. 

% %\subsubsection{Quantitative Analysis}
% Direct evaluation of Style Quotient is difficult as there is no equivalent ground truth available which rates the stylistic content. So, we assess the performance of SQ based demand model on its prediction accuracy for test data and compare the same with various baseline models. Comparison is done for following models - 
% \begin{enumerate}
% \item Simple ROS Model - 
% \item Normalize ROS Model - 
% \item Mean Intercept Model - 
% \item SQ based Model - 
% \end{enumerate}
% \fi

%% file: sample-sigconf-authordraft.bbl
%%% -*-BibTeX-*-
%%% Do NOT edit. File created by BibTeX with style
%%% ACM-Reference-Format-Journals [18-Jan-2012].

\begin{thebibliography}{00}

%%% ====================================================================
%%% NOTE TO THE USER: you can override these defaults by providing
%%% customized versions of any of these macros before the \bibliography
%%% command.  Each of them MUST provide its own final punctuation,
%%% except for \shownote{}, \showDOI{}, and \showURL{}.  The latter two
%%% do not use final punctuation, in order to avoid confusing it with
%%% the Web address.
%%%
%%% To suppress output of a particular field, define its macro to expand
%%% to an empty string, or better, \unskip, like this:
%%%
%%% \newcommand{\showDOI}[1]{\unskip}   % LaTeX syntax
%%%
%%% \def \showDOI #1{\unskip}           % plain TeX syntax
%%%
%%% ====================================================================

\ifx \showCODEN    \undefined \def \showCODEN     #1{\unskip}     \fi
\ifx \showDOI      \undefined \def \showDOI       #1{#1}\fi
\ifx \showISBNx    \undefined \def \showISBNx     #1{\unskip}     \fi
\ifx \showISBNxiii \undefined \def \showISBNxiii  #1{\unskip}     \fi
\ifx \showISSN     \undefined \def \showISSN      #1{\unskip}     \fi
\ifx \showLCCN     \undefined \def \showLCCN      #1{\unskip}     \fi
\ifx \shownote     \undefined \def \shownote      #1{#1}          \fi
\ifx \showarticletitle \undefined \def \showarticletitle #1{#1}   \fi
\ifx \showURL      \undefined \def \showURL       {\relax}        \fi
% The following commands are used for tagged output and should be
% invisible to TeX
\providecommand\bibfield[2]{#2}
\providecommand\bibinfo[2]{#2}
\providecommand\natexlab[1]{#1}
\providecommand\showeprint[2][]{arXiv:#2}

\bibitem[\protect\citeauthoryear{Bracher, Heinz, and Vollgraf}{Bracher
  et~al\mbox{.}}{2016}]%
        {bracher2016fashion}
\bibfield{author}{\bibinfo{person}{Christian Bracher},
  \bibinfo{person}{Sebastian Heinz}, {and} \bibinfo{person}{Roland Vollgraf}.}
  \bibinfo{year}{2016}\natexlab{}.
\newblock \showarticletitle{Fashion DNA: Merging content and sales data for
  recommendation and article mapping}.
\newblock \bibinfo{journal}{{\em arXiv preprint arXiv:1609.02489\/}}
  (\bibinfo{year}{2016}).
\newblock


\bibitem[\protect\citeauthoryear{Cachon, Terwiesch, and Xu}{Cachon
  et~al\mbox{.}}{2005}]%
        {cachon2005retail}
\bibfield{author}{\bibinfo{person}{G{\'e}rard~P Cachon},
  \bibinfo{person}{Christian Terwiesch}, {and} \bibinfo{person}{Yi Xu}.}
  \bibinfo{year}{2005}\natexlab{}.
\newblock \showarticletitle{Retail assortment planning in the presence of
  consumer search}.
\newblock \bibinfo{journal}{{\em Manufacturing \& Service Operations
  Management\/}} \bibinfo{volume}{7}, \bibinfo{number}{4}
  (\bibinfo{year}{2005}), \bibinfo{pages}{330--346}.
\newblock


\bibitem[\protect\citeauthoryear{Caro and Gallien}{Caro and Gallien}{2007}]%
        {caro2007dynamic}
\bibfield{author}{\bibinfo{person}{Felipe Caro} {and}
  \bibinfo{person}{J{\'e}r{\'e}mie Gallien}.} \bibinfo{year}{2007}\natexlab{}.
\newblock \showarticletitle{Dynamic assortment with demand learning for
  seasonal consumer goods}.
\newblock \bibinfo{journal}{{\em Management Science\/}} \bibinfo{volume}{53},
  \bibinfo{number}{2} (\bibinfo{year}{2007}), \bibinfo{pages}{276--292}.
\newblock


\bibitem[\protect\citeauthoryear{Cooper and Nakanishi}{Cooper and
  Nakanishi}{1988}]%
        {cooper1988market}
\bibfield{author}{\bibinfo{person}{Lee~G Cooper} {and} \bibinfo{person}{Masao
  Nakanishi}.} \bibinfo{year}{1988}\natexlab{}.
\newblock \bibinfo{booktitle}{{\em Market-Share Analvsis. Evaluatinq
  Comoetitive Marketing Effectiveness. International Series in Quantitative
  Marketing (ISQM)}}.
\newblock \bibinfo{publisher}{Boston, MA: Kluwer Academic Publishers}.
\newblock


\bibitem[\protect\citeauthoryear{Garg, Banerjee, Anoop, Sreenivas, and
  Warrier}{Garg et~al\mbox{.}}{2016}]%
        {garg2016sales}
\bibfield{author}{\bibinfo{person}{Vikram Garg}, \bibinfo{person}{Rajdeep~H
  Banerjee}, \bibinfo{person}{KR Anoop}, \bibinfo{person}{T Sreenivas}, {and}
  \bibinfo{person}{Deepak Warrier}.} \bibinfo{year}{2016}\natexlab{}.
\newblock \showarticletitle{Sales Potential: Modelling Sellability of Visual
  Aesthetics of a Fashion Product}.
\newblock  (\bibinfo{year}{2016}).
\newblock


\bibitem[\protect\citeauthoryear{Gaur and Honhon}{Gaur and Honhon}{2006}]%
        {gaur2006assortment}
\bibfield{author}{\bibinfo{person}{Vishal Gaur} {and}
  \bibinfo{person}{Doroth{\'e}e Honhon}.} \bibinfo{year}{2006}\natexlab{}.
\newblock \showarticletitle{Assortment planning and inventory decisions under a
  locational choice model}.
\newblock \bibinfo{journal}{{\em Management Science\/}} \bibinfo{volume}{52},
  \bibinfo{number}{10} (\bibinfo{year}{2006}), \bibinfo{pages}{1528--1543}.
\newblock


\bibitem[\protect\citeauthoryear{K{\"o}k and Fisher}{K{\"o}k and
  Fisher}{2007}]%
        {kok2007demand}
\bibfield{author}{\bibinfo{person}{A~G{\"u}rhan K{\"o}k} {and}
  \bibinfo{person}{Marshall~L Fisher}.} \bibinfo{year}{2007}\natexlab{}.
\newblock \showarticletitle{Demand estimation and assortment optimization under
  substitution: Methodology and application}.
\newblock \bibinfo{journal}{{\em Operations Research\/}} \bibinfo{volume}{55},
  \bibinfo{number}{6} (\bibinfo{year}{2007}), \bibinfo{pages}{1001--1021}.
\newblock


\bibitem[\protect\citeauthoryear{K{\"o}k, Fisher, and Vaidyanathan}{K{\"o}k
  et~al\mbox{.}}{2008}]%
        {kok2008assortment}
\bibfield{author}{\bibinfo{person}{A~G{\"u}rhan K{\"o}k},
  \bibinfo{person}{Marshall~L Fisher}, {and} \bibinfo{person}{Ramnath
  Vaidyanathan}.} \bibinfo{year}{2008}\natexlab{}.
\newblock \showarticletitle{Assortment planning: Review of literature and
  industry practice}.
\newblock In \bibinfo{booktitle}{{\em Retail supply chain management}}.
  \bibinfo{publisher}{Springer}, \bibinfo{pages}{99--153}.
\newblock


\bibitem[\protect\citeauthoryear{Maddah and Bish}{Maddah and Bish}{2007}]%
        {maddah2007joint}
\bibfield{author}{\bibinfo{person}{Bacel Maddah} {and} \bibinfo{person}{Ebru~K
  Bish}.} \bibinfo{year}{2007}\natexlab{}.
\newblock \showarticletitle{Joint pricing, assortment, and inventory decisions
  for a retailer's product line}.
\newblock \bibinfo{journal}{{\em Naval Research Logistics (NRL)\/}}
  \bibinfo{volume}{54}, \bibinfo{number}{3} (\bibinfo{year}{2007}),
  \bibinfo{pages}{315--330}.
\newblock


\bibitem[\protect\citeauthoryear{Mahajan and Van~Ryzin}{Mahajan and
  Van~Ryzin}{2001}]%
        {mahajan2001stocking}
\bibfield{author}{\bibinfo{person}{Siddharth Mahajan} {and}
  \bibinfo{person}{Garrett Van~Ryzin}.} \bibinfo{year}{2001}\natexlab{}.
\newblock \showarticletitle{Stocking retail assortments under dynamic consumer
  substitution}.
\newblock \bibinfo{journal}{{\em Operations Research\/}} \bibinfo{volume}{49},
  \bibinfo{number}{3} (\bibinfo{year}{2001}), \bibinfo{pages}{334--351}.
\newblock


\bibitem[\protect\citeauthoryear{Ryzin and Mahajan}{Ryzin and Mahajan}{1999}]%
        {ryzin1999relationship}
\bibfield{author}{\bibinfo{person}{Garrett~van Ryzin} {and}
  \bibinfo{person}{Siddharth Mahajan}.} \bibinfo{year}{1999}\natexlab{}.
\newblock \showarticletitle{On the relationship between inventory costs and
  variety benefits in retail assortments}.
\newblock \bibinfo{journal}{{\em Management Science\/}} \bibinfo{volume}{45},
  \bibinfo{number}{11} (\bibinfo{year}{1999}), \bibinfo{pages}{1496--1509}.
\newblock


\bibitem[\protect\citeauthoryear{Smith and Agrawal}{Smith and Agrawal}{2000}]%
        {smith2000management}
\bibfield{author}{\bibinfo{person}{Stephen~A Smith} {and}
  \bibinfo{person}{Narendra Agrawal}.} \bibinfo{year}{2000}\natexlab{}.
\newblock \showarticletitle{Management of multi-item retail inventory systems
  with demand substitution}.
\newblock \bibinfo{journal}{{\em Operations Research\/}} \bibinfo{volume}{48},
  \bibinfo{number}{1} (\bibinfo{year}{2000}), \bibinfo{pages}{50--64}.
\newblock


\bibitem[\protect\citeauthoryear{Wang, Nabi, Wang, Wan, and Ng}{Wang
  et~al\mbox{.}}{2015}]%
        {wang2015towards}
\bibfield{author}{\bibinfo{person}{Jinghua Wang}, \bibinfo{person}{Abrar~Abdul
  Nabi}, \bibinfo{person}{Gang Wang}, \bibinfo{person}{Chengde Wan}, {and}
  \bibinfo{person}{Tian-Tsong Ng}.} \bibinfo{year}{2015}\natexlab{}.
\newblock \showarticletitle{Towards Predicting the Likeability of Fashion
  Images}.
\newblock \bibinfo{journal}{{\em arXiv preprint arXiv:1511.05296\/}}
  (\bibinfo{year}{2015}).
\newblock


\end{thebibliography}
